\documentclass[reprint,amsmath,amssymb,aps,prc,superscriptaddress]{revtex4-2}
\usepackage[utf8]{inputenc}

\usepackage{graphicx}
\usepackage{dcolumn}
\usepackage{bm}
\usepackage{hyperref}

\usepackage{pgfplots}
\DeclareUnicodeCharacter{2212}{−}
\usepgfplotslibrary{groupplots,dateplot}
\usetikzlibrary{patterns,shapes.arrows}
\pgfplotsset{compat=newest}

\pgfmathdeclarefunction{lg10}{1}{%
    \pgfmathparse{ln(#1)/ln(10)}%
}

\pgfplotsset{error bar legend/.style={%
    /pgfplots/legend image code/.prefix code={%
      \pgfkeysgetvalue{/pgfplots/error bars/error mark}{\pgfplotserrorbarsmark}%
      \draw[%
        /pgfplots/every error bar, 
        mark=\pgfplotserrorbarsmark, 
        /pgfplots/error bars/error mark options, 
        sharp plot,
        ##1
      ] plot coordinates {(0.3cm, -0.15cm) (0.3cm, 0.15cm)};%
    }
  }
}

\begin{document}

\preprint{APS/PRC}

\title{\texorpdfstring{Nuclear Level Density and $\gamma$-ray Strength Function of $^{63}\mathrm{Ni}$}{Nuclear Level Density and Gamma-ray Strength Function of 63-Ni}}

\author{V.~W.~Ingeberg}
\affiliation{Department of Physics, University of Oslo, N-0316 Oslo, Norway}

\author{P.~Jones}
\affiliation{Department of Subatomic Physics, iThemba LABS, P.O. Box 722, 7129 Somerset West, South Africa}

\author{L.~Msebi}
\affiliation{Department of Subatomic Physics, iThemba LABS, P.O. Box 722, 7129 Somerset West, South Africa}
\affiliation{Physics Department, University of the Western Cape, P/B X17, Bellville, 7535, South Africa}

\author{S.~Siem}
\affiliation{Department of Physics, University of Oslo, N-0316 Oslo, Norway}

\author{M.~Wiedeking}
\affiliation{Department of Subatomic Physics, iThemba LABS, P.O. Box 722, 7129 Somerset West, South Africa}
\affiliation{School of Physics, University of the Witwatersrand, 2050 Johannesburg, South Africa}

\author{A.~A.~Avaa}
\affiliation{Department of Subatomic Physics, iThemba LABS, P.O. Box 722, 7129 Somerset West, South Africa}
\affiliation{Physics Department, University of the Western Cape, P/B X17, Bellville, 7535, South Africa}

\author{M.~V.~Chisapi}
\affiliation{Department of Subatomic Physics, iThemba LABS, P.O. Box 722, 7129 Somerset West, South Africa}
\affiliation{Physics Department, University of the Western Cape, P/B X17, Bellville, 7535, South Africa}
\affiliation{Physics Department, Stellenbosch University, P/B X1, Matieland, 7602, South Africa}

\author{E.~A.~Lawrie}
\affiliation{Department of Subatomic Physics, iThemba LABS, P.O. Box 722, 7129 Somerset West, South Africa}

\author{K.~L.~Malatji}
\affiliation{Department of Subatomic Physics, iThemba LABS, P.O. Box 722, 7129 Somerset West, South Africa}
\affiliation{Physics Department, Stellenbosch University, P/B X1, Matieland, 7602, South Africa}

\author{L.~Makhathini}
\affiliation{Department of Subatomic Physics, iThemba LABS, P.O. Box 722, 7129 Somerset West, South Africa}

\author{S.~P.~Noncolela}
\affiliation{Department of Subatomic Physics, iThemba LABS, P.O. Box 722, 7129 Somerset West, South Africa}
\affiliation{Physics Department, University of the Western Cape, P/B X17, Bellville, 7535, South Africa}

\author{O.~Shirinda}
\affiliation{Department of Physical and Earth Sciences, Sol Plaatje University, Private Bag X5008, Kimberley 8301, South Africa}

\date{\today}

\begin{abstract}
    The nuclear level density (NLD) and $\gamma$-ray strength function ($\gamma$SF) of $^{63}\mathrm{Ni}$ have been investigated using the Oslo method. The extracted NLD is compared with previous measurements using particle evaporation \cite{Voinov2012} and those found from neutron resonance spacing \cite{mughabghab2018atlas,Capote2009,PhysRevC.89.025810}. The $\gamma$SF was found to feature a strong low energy enhancement that could be explained as M1 strength based on large scale shell model calculations \cite{Midtbo2018}. Comparison of $\gamma$SFs measured with the Oslo method for various $\mathrm{Ni}$ isotopes reveals systematic changes to the strength below $5$ MeV with increasing mass.
\end{abstract}

\maketitle

\section{Introduction}
The Oslo method is a powerful analytical method that allows for simultaneous extraction of nuclear level density (NLD) and $\gamma$-ray strength functions ($\gamma$SF) from particle-$\gamma$ coincidences following reactions with light ion beams (e.g. $(p,p^\prime)$, $(d,p)$ etc.) \cite{OsloMethodNIM}. The method has been extended to be used in conjunction with total absorption spectrometry following $\beta$-decay ($\beta$-Oslo method) \cite{PhysRevLett.113.232502} and particle-$\gamma$ coincidences from inverse-kinematics experiments \cite{Ingeberg2020}. 

The Oslo method itself does not provide the absolute NLD and $\gamma$SF values, but rather the functional shapes. In order to determine the correct common slope of the NLD and $\gamma$SF, as well as their absolute values, a normalization to auxiliary experimental data is required. Typical data for normalization are the s-wave resonance spacing, discrete resolved levels and average radiative width. The reliance on external data means that the accuracy of the final NLD and $\gamma$SF is mostly determined by the accuracy of those data. The resonance spacings and radiative widths can be highly uncertain, especially in nuclei with few resonances. For the majority of unstable nuclei these have not even been measured. This means that alternative approaches for normalization have to be used especially for cases where no experimental resonance data are available. For nuclei close to stability these values can typically be estimated from systematics in the vicinity of the nucleus using models \cite{PhysRevLett.113.232502}. The down side of such normalized NLDs and $\gamma$SFs is the introduction of model dependencies which may result in large uncertainties. A model independent approach is the use of the Shape method \cite{wiedeking2020independent,mucher2020novel} to determine the slope of the $\gamma$SF, however the method requires sufficient particle energy resolution and a well known level structure with resolvable energy spacing at low excitation energy. In this paper we will look at a possible third option in which only NLD from known discrete states is used to normalize the NLD.

In this paper we have analyzed data from a $(\mathrm{p},\mathrm{d})$ reaction on $^{64}\mathrm{Ni}$ to measure the NLD and $\gamma$SF of $^{63}\mathrm{Ni}$. The level density of $^{63}\mathrm{Ni}$ has previously been measured using particle evaporation spectra and shows significantly lower NLD than that expected from resonance spacing data \cite{PhysRevC.80.034305,Voinov2012}. This makes $^{63}\mathrm{Ni}$ a very interesting case study as a normalization only considering known discrete levels could resolve the discrepancy. In addition, the $\gamma$SFs have previously been measured in several other $\mathrm{Ni}$ isotopes and consistently show a strong low energy enhancement \cite{PhysRevC.94.044321,PhysRevC.96.014312,PhysRevC.98.054619,PhysRevC.97.054329,Spyrou2017,Ingeberg2022a}. With this measurement the NLD and $\gamma$SF will have been measured in most stable \cite{Renstrom2018,PhysRevC.94.044321,PhysRevC.96.014312,PhysRevC.98.054619} and several unstable $\mathrm{Ni}$ isotopes \cite{Renstrom2018,PhysRevC.97.054329,Spyrou2017,Ingeberg2022a}, allowing for investigations into the systematics of the $\gamma$SF.

\section{Experiment and analysis}
The experiment measuring particle-$\gamma$ coincidences from the $^{64}\mathrm{Ni}(\mathrm{p},\mathrm{d})^{63}\mathrm{Ni}$ reaction was performed with a $27.4$ MeV proton beam accelerated by the Separated Sector Cyclotron (SSC) at iThemba LABS. The $4.56$ mg/cm$^2$ thick $^{64}\mathrm{Ni}$ target was bombarded with a beam current of $\approx 1$ pnA for about $15$ hours at the center of the AFRODITE array \cite{NIMA2007}. The array consisted of eight Compton suppressed high purity germanium (HPGe) CLOVER detectors, six small (2"x2") and two large volume (3.5"x8") LaBr$_3$:Ce detectors. Particles from the reaction were measured by two silicon detectors of the S2 type in a $\Delta$E-E configuration and placed down stream of the target. The $\Delta$E detector had a thickness of $309$ $\mu$m and the E detector was $1041$ $\mu$m thick. In front of the particle telescope a $10$ $\mu$m thick aluminum absorber was placed to shield from $\delta$-electrons. Signals from the detectors were read out using Pixie-16 digital pulse processors from XIA. Each detector was self-triggering and the pulse height, timestamp and constant fraction corrections of each event were stored to disk for offline analysis.

\begin{figure*}
    \centering
    \includegraphics[width=\textwidth]{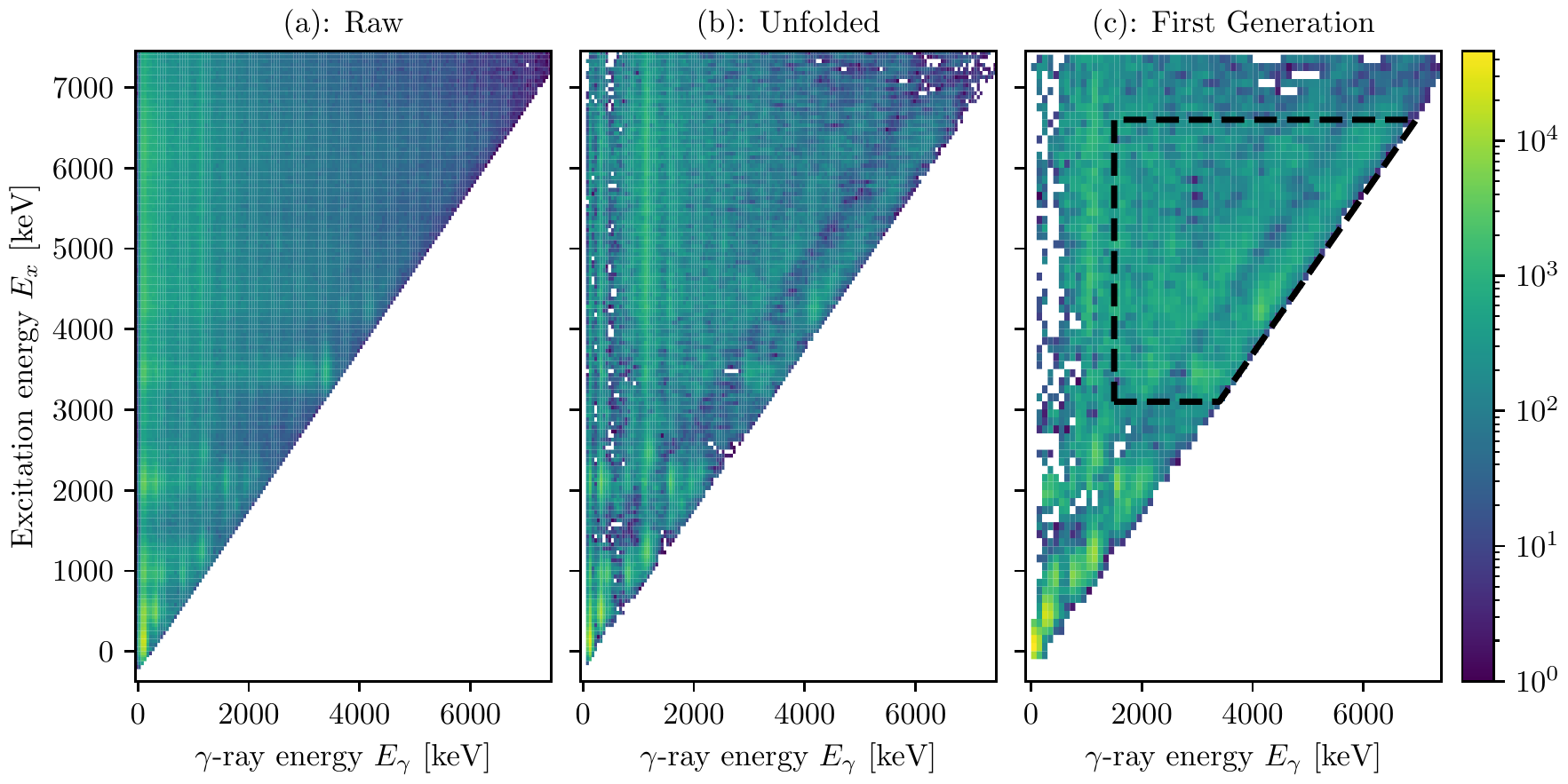}
    \caption{The (a) raw, (b) unfolded and (c) first generation matrix.}
    \label{fig:matrices}
\end{figure*}

Particle-$\gamma$ coincidences were found in the list mode data by placing time gates on the prompt time peak in the particle-$\gamma$ time spectra. Background events were found by placing an off-prompt time gate of similar length. The mass and charge of the ejected particle, and thus the reaction channel, was selected by applying a graphical cut in the $\Delta\mathrm{E}$ vs. $\mathrm{E}$ matrix. For each event the excitation energy of the residual $^{63}\mathrm{Ni}$ nucleus was found from kinematic reconstruction assuming a two-body reaction. The resulting excitation energy and coincident $\gamma$-ray spectrum were then used to construct the prompt excitation versus $\gamma$-ray energy matrix shown in Fig. \ref{fig:matrices}(a). A similar background excitation-$\gamma$-ray energy matrix was constructed from the events in the background time gate. After applying time and particle gates a total of $3.7\times10^6$, $4.8\times10^6$ and $5.7\times10^6$  prompt particle-$\gamma$ coincidences and $7.3\times10^5$, $1.0\times10^6$ and $7.9\times10^5$ background events were found in the CLOVER, large LaBr$_3$:Ce and small LaBr$_3$:Ce detectors, respectively. The considerably lower background to prompt ratio for the small LaBr$_3$:Ce detectors can be attributed to their exceptionally high time resolution \cite{MSEBI2022166195}. In the following analysis only particle-$\gamma$ coincidences in the large LaBr$_3$:Ce detectors were considered as these exhibit far superior efficiency at high $\gamma$-ray energies which is important in the Oslo method.

\subsection{The Oslo method}\label{subsec:OsloMethod}
The starting point for the Oslo method is the excitation-$\gamma$ matrix. The first step is to correct for the response of the $\gamma$-detector using the \emph{unfolding method} \cite{UnfoldingNIM}. The response function of the setup was found from simulations of the AFRODITE array using a model implemented in \verb+Geant4+ \cite{AGOSTINELLI2003250,Ingeberg2022}. The resulting unfolded matrix is shown in Fig. \ref{fig:matrices}(b). The peak at $E_x = 3.6$ MeV to the ground state was fitted and subtracted from the unfolded spectra with the justification being that this state is only populated directly from the reaction and has no feeding from the quasi-continuum.

Next is to find the first generation matrix using the \emph{first generation method} \cite{FirstGenerationNIM}. The resulting first generation matrix contains the distribution of the first $\gamma$-rays emitted in cascades depopulating each excitation bin and is shown in Fig. \ref{fig:matrices}(c).

The first generation matrix is proportional to the NLD and $\gamma$-ray transmission coefficient via \cite{OsloMethodNIM}
\begin{equation}
    \Gamma(E_\gamma, E_x) \propto \mathcal{T}(E_\gamma) \rho(E_x - E_\gamma), \label{eq:FGprop}
\end{equation}
where $\Gamma(E_\gamma, E_x)$ is the bin with $\gamma$-ray energy $E_\gamma$ and excitation energy $E_x$. $\mathcal{T}(E_\gamma)$ is the transmission coefficient for $\gamma$-ray energy $E_\gamma$ and $\rho(E_x - E_\gamma)$ is the level density at the final excitation energy $E_f = E_x - E_\gamma$. The NLD and $\gamma$-ray transmission coefficients are extracted from the first generation matrix by fitting a theoretical matrix
\begin{equation}
    \Gamma_\text{th}(E_\gamma, E_x) = \frac{\rho(E_x - E_\gamma) \mathcal{T}(E_\gamma)}{\sum\limits_{ E_\gamma = E_\gamma^\text{min} }^{E_x} \rho(E_x-E_\gamma) \mathcal{T}(E_\gamma)}, \label{eq:FGtheo}
\end{equation}
where $\rho(E_x - E_\gamma)$ and $\mathcal{T}(E_\gamma)$ are treated as free variables for each final energy $E_f = E_x - E_\gamma$ and $\gamma$-ray energy $E_\gamma$. The fit was done by minimizing
\begin{equation}
    \chi^2 = \sum\limits_{E_x, E_\gamma} \left( \frac{\Gamma(E_\gamma, E_x) - \Gamma_\text{th}(E_\gamma, E_x)}{\Delta \Gamma(E_\gamma, E_x)} \right)^2. \label{eq:ChiSqMin}
\end{equation}
The region of the first generation matrix fitted was limited to a minimum $\gamma$-ray energy of $1500$ keV and excitation energies between $3100$ keV and $6600$ keV to ensure only statistical decay was included. The region is highlighted by the dashed line in Fig. \ref{fig:matrices}(c).

\begin{figure}
    \centering
    \includegraphics[width=0.5\textwidth]{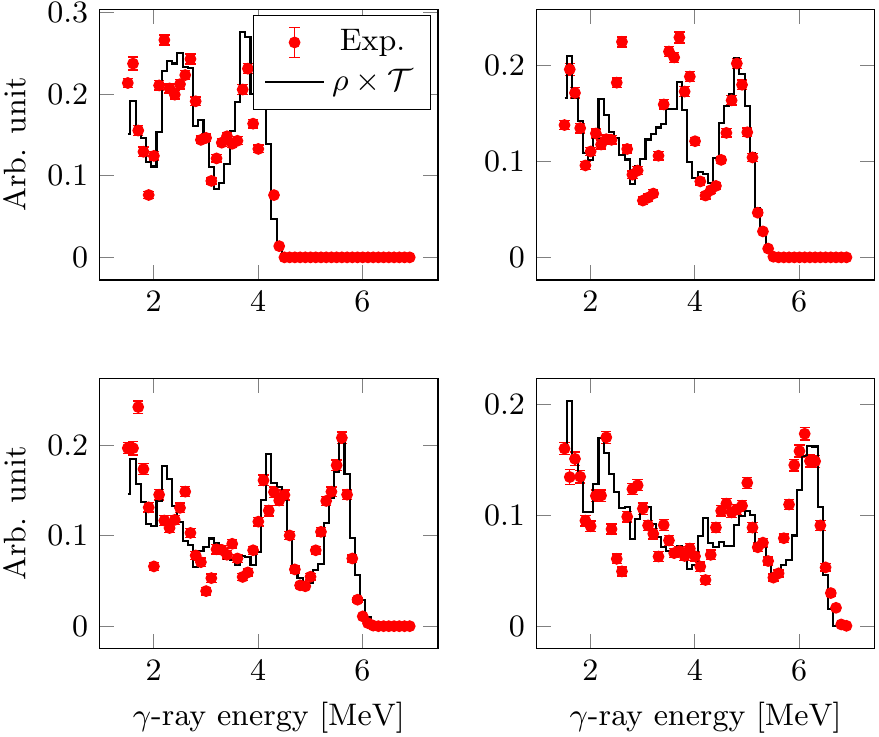}
    \caption{$^{63}\mathrm{Ni}$ primary $\gamma$-ray distribution at excitation energy $4$ MeV (upper left), $5$ MeV (upper right), $5.75$ MeV (lower left) and $6.4$ MeV (lower right). Red dots show the experimental first generation spectra, while the solid black line is the product of the fitted NLD and $\gamma$SF.}
    \label{fig:does_it_work}
\end{figure}

The resulting theoretical first generation matrix are shown for a few select excitation bins together with the experimental matrix in Fig. \ref{fig:does_it_work}.

The NLD and $\gamma$-ray transmission coefficients resulting from the $\chi^2$ minimization are not the physical values, but rather the shape as Eq. \eqref{eq:ChiSqMin} is symmetric under transformation
\begin{equation}
    \begin{split}
        \tilde{\rho}(E_x - E_\gamma) &= A \rho(E_x - E_\gamma) e^{\alpha(E_x - E_\gamma)} \\
        \tilde{\mathcal{T}}(E_\gamma) &= B \mathcal{T}(E_\gamma) e^{\alpha E_\gamma}, \label{eq:NLD_T_transform}
    \end{split}
\end{equation}
where $A$, $B$ and $\alpha$ are transformation parameters. To obtain the physical transformation for the extracted NLD and $\gamma$-transmission coefficients, a normalization to external data has to be performed, see Sect. \ref{sec:Normalization}. The $\gamma$SF is related to the transmission coefficient via $f(E_\gamma) = \mathcal{T}(E_\gamma)/(2\pi E_\gamma^3)$, under the assumption that dipole transitions dominate the transmission coefficients.

\section{\texorpdfstring{Normalization of level density \& $\gamma$-ray strength function}{Normalization of level density \& gamma-ray strength function}}\label{sec:Normalization}

The main auxiliary data required to normalize the NLD is known level densities from tabulated levels and the NLD at the neutron separation energy $S_n$. Tabulated levels are converted to level density simply by counting the number of levels within each excitation bin and dividing by the bin width. This results in a level density that will have large fluctuations compared to Oslo method data as the experimental resolution has not yet been accounted for. The level density from known levels is smoothed with a Gaussian with FWHM of about $325$ keV to match the experimental resolution for final excitation energy. Tabulated discrete levels were taken from the RIPL-3 library \cite{Capote2009}.

The level density at the neutron separation energy is found from the resonance spacing of s-wave resonances $D_0$ by \cite{OsloMethodNIM}
\begin{equation}
    \rho(S_n) = \frac{2}{g(S_n, J_t-1/2) + g(S_n, J_t+1/2)}\frac{1}{D_0} \label{eq:NLDSn}
\end{equation}
where $J_t$ is the ground state spin of the $A-1$ nucleus. The spin distribution $g(E_x, J)$ is given by the Ericson distribution \cite{ERICSON1959481}
\begin{equation}
    g(E_x, J) = \exp\left( -\frac{J^2}{2\sigma^2(E_x)}\right) - \exp\left( -\frac{(J + 1)^2}{2\sigma^2(E_x)} \right), \label{eq:EricsonDist}
\end{equation}
with the spin-cutoff parameter parameterized by \cite{PhysRevC.96.024313}
\begin{equation}
    \sigma^2(E_x) = \begin{cases}
    \sigma_d^2 & \quad E < E_d \\
    \sigma_d^2 + \frac{E - E_d}{S_n - E_d}(\sigma^2(S_n) - \sigma^2_d) & \quad E \geq E_d.
    \end{cases}
    \label{eq:SpinCutParam}
\end{equation}
The spin-cutoff parameter of the discrete levels ($E_d = 2.0$ MeV) is estimated to be $\sigma_d = 2.30(23)$ from tabulated discrete levels \cite{Capote2009} and large scale shell model calculations \cite{Midtbo2018}, while the spin-cutoff parameter at the neutron separation energy was estimated to be $\sigma(S_n) = 3.68(21)$ estimated from the models of refs. \cite{VonEgidy2005}, \cite{doi:10.1139/p65-139} and \cite{PhysRevC.80.054310}. The s-wave resonance spacing $D_0=16.0(30)$ keV was taken from the RIPL-3 database \cite{Capote2009} resulting in a total level density at the neutron separation energy of $1730(363)$ MeV$^{-1}$. 

The experimental NLD only extends up to $5.2$ MeV and to properly compare with the level density at the neutron separation energy the NLD is extrapolated to $S_n$ via a constant temperature (CT) formula \cite{ERICSON1959481}
\begin{equation}
    \rho_\text{CT}(E_x) = \frac{1}{T}\exp\left(\frac{E_x - E_\text{shift}}{T}\right), \label{eq:ConstTemp}
\end{equation}
where the temperature $T$ and shift parameter $E_\text{shift}$ are treated as free parameters. 

\begin{figure}
    \centering
    \includegraphics[width=0.5\textwidth]{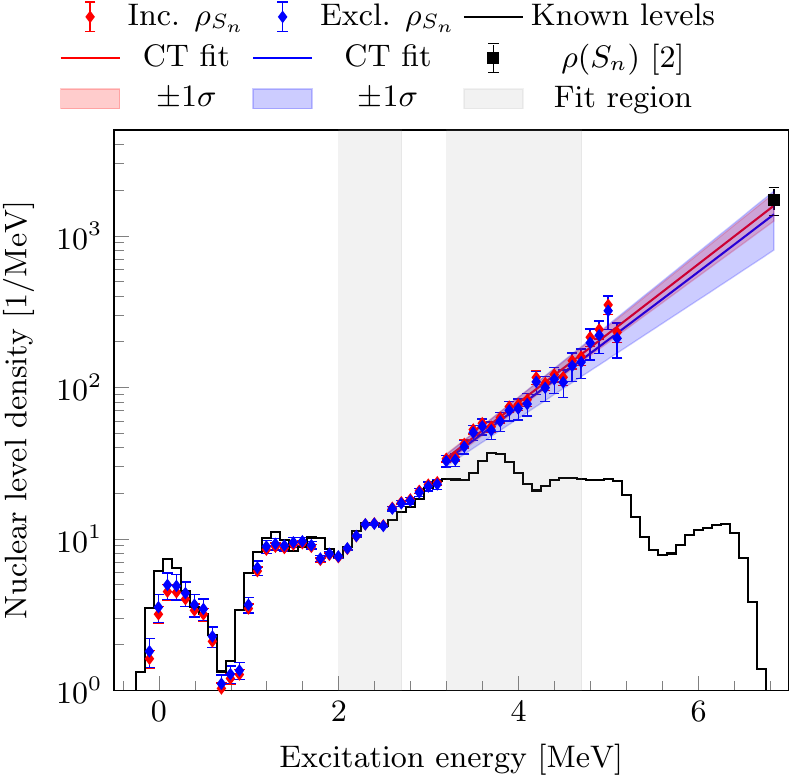}
    \caption{The extracted and normalized NLD. The red and blue circles are the experimental values, while the black solid line is the level density from the known resolved levels convoluted with the experimental resolution. The black dash-dotted line is the level density found from large scale shell model calculations using the ca48mh1 interaction \cite{Midtbo2018}. The black solid square is the level density at the neutron separation energy found from s-wave resonance spacing reported by \cite{PhysRevC.89.025810,Capote2009,mughabghab2018atlas}. The red solid line shows the level density from the fitted CT model while the red shaded area is the $\pm1\sigma$ confidence interval.}
    \label{fig:nld_conv}
\end{figure}

\begin{figure}
    \centering
    \includegraphics[width=0.5\textwidth]{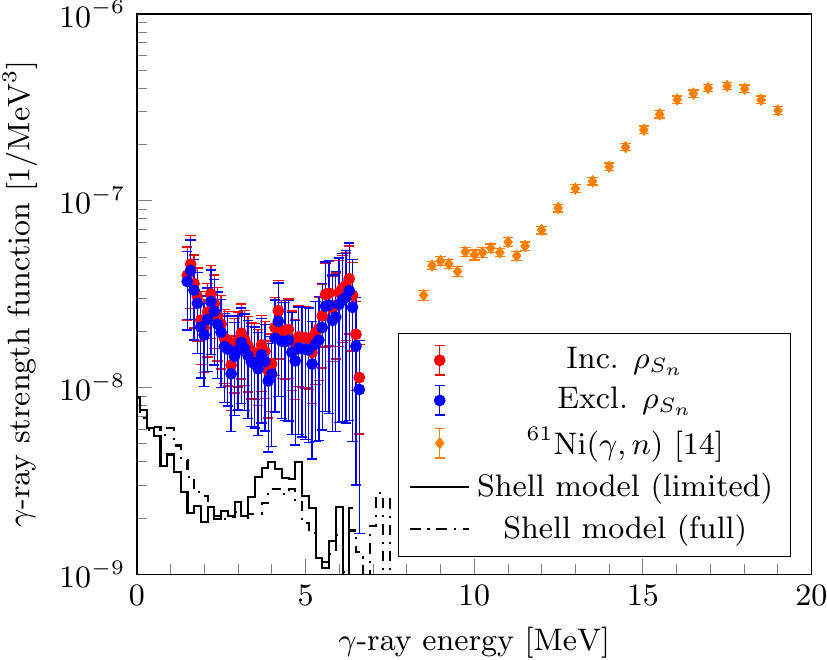}
    \caption{Extracted $\gamma$SF when including the NLD at $S_n$ from resonance spacings in the normalization are shown by the red circles while the blue circles only considers the level density from known levels. The orange diamonds are the $\gamma$SF of $^{61}\mathrm{Ni}$ measured by \cite{PhysRevC.98.054619}. The black line shows the calculated M1 strength from shell model calculations \cite{Midtbo2018} considering only decay from levels within the fit region, while the dash-dotted line includes all levels found in the shell model calculation.}
    \label{fig:gsf_conv}
\end{figure}

Data required to normalize the $\gamma$SF is the average radiative width of s-wave resonances, as this value is related to the $\gamma$SF and NLD via \cite{PhysRevC.41.1941}
\begin{equation}
    \begin{split}
    \langle \Gamma_{\gamma0} \rangle = \frac{D_0}{2} \int_0^{S_n} dE_\gamma E_\gamma^3 f(E_\gamma) \rho(S_n - E_\gamma) \\
    \times  \left[g(S_n - E_\gamma, 1/2) + g(S_n - E_\gamma, 3/2)\right] E_\gamma.
\end{split}
\label{eq:RadWidth}
\end{equation}
Due to the limits selected (see Sect. \ref{subsec:OsloMethod}) for the extraction of the NLD and $\gamma$SF the experimental data only extends up to $E_x=5.2$ MeV and $E_\gamma$ between $1.5$ and $6.6$ MeV, respectively. To evaluate the integral in eq. \eqref{eq:RadWidth} the NLD was extrapolated with the constant temperature formula, eq. \eqref{eq:ConstTemp}, between $5.2$ MeV and the neturon separation energy. The $\gamma$SF was extrapolated using $f(E_\gamma) = Ce^{\eta E_\gamma}$, and $f(E_\gamma) = Ce^{\eta E_\gamma}/E_\gamma^3$ for energies between $0$ and $1.5$ MeV, and $6.6$ MeV and the neutron separation energy, respectively. The average radiative width of s-wave resonances in $^{63}\mathrm{Ni}$ was found to be $534(214)$ meV by a weighted average of all the tabulated values found in \cite{PhysRevC.89.025810}. Due to the large spread of the tabulated values a large uncertainty of $40\%$ was assumed. All normalization parameters adopted in this analysis are listed in Table \ref{tab:param_table}. The normalization parameters $A$, $B$ and $\alpha$ were found by sampling the posterior probability distribution with total likelihood function
\begin{equation}
    \mathcal{L}(\bm{\theta}) = \prod_i \mathcal{L}_i(\bm{\theta}). \label{eq:tot_likelihood}
\end{equation}
using the Bayesian sampling package \verb+UltraNest+ \cite{2021JOSS6.3001B}. All experimental data are assumed to be normally distributed, giving the likelihoods
\begin{align}
\begin{split}
    \ln \mathcal{L}_\text{discrete} &= \sum_i \ln\frac{1}{\sqrt{2\pi \sigma_{j,\text{Oslo}}(\bm{\theta})}} \\
    &- \frac{1}{2}\sum_i \left(\frac{\rho_{j,\text{discrete}} - \rho_{j,\text{Oslo}}(\bm{\theta})}{\sigma_{j,\text{Oslo}}(\bm{\theta})} \right)^2, \label{eq:discrete_like}
\end{split} \\
\begin{split}
    \ln \mathcal{L}_\text{CT} &= \sum_i \ln\frac{1}{\sqrt{2\pi \sigma_{j,\text{Oslo}}(\bm{\theta})}} \\
    &- \frac{1}{2}\sum_i \left(\frac{\rho_{j,\text{CT}} - \rho_{j,\text{Oslo}}(\bm{\theta})}{\sigma_{j,\text{Oslo}}(\bm{\theta})} \right)^2, \label{eq:CT_like}
\end{split} \\
\ln \mathcal{L}_{\rho_{S_n}} &= \left( \frac{\rho_{S_n} - \rho_{S_n,\text{CT}}(\bm{\theta})}{\sigma_{\rho_{S_n}}} \right)^2,  \label{eq:rhoSn_like} \\
\ln \mathcal{L}_{\langle \Gamma_{\gamma0} \rangle} &= \left( \frac{\langle\Gamma_{\gamma0} \rangle_\text{exp} - \langle \Gamma_{\gamma0} \rangle_\text{Oslo}(\bm{\theta})}{\sigma_{\langle\Gamma_{\gamma0} \rangle_\text{exp}}} \right)^2. \label{eq:Gg0_like}
\end{align}
The parameters $\bm{\theta} = (A, B, \alpha, T, E_\text{shift}, \sigma_D, \sigma_{S_n})$ have a uniform prior between $0$ and $5$ for $A$ and $B$ and $-1$ MeV$^{-1}$ and $1$ MeV$^{-1}$ for $\alpha$. The temperature and shift parameters also used a uniform prior between $0.2$ and $2$ MeV and $-10$ and $10$ MeV, respectively. The spin cut-off parameters were included as nuance parameters with normal distributed priors to ensure proper propagation of errors. The resulting normalized NLD and $\gamma$SF are shown as red circles in figs. \ref{fig:nld_conv} and \ref{fig:gsf_conv}, respectively. The discrete likelihood, eq. \eqref{eq:discrete_like} was limited to data points between $2$ and $2.7$ MeV, while the CT formula was fitted between $3.2$ and $4.7$ MeV. To investigate the sensitivity to the resonance spacing the analysis was repeated, but excluding Eq. \eqref{eq:rhoSn_like} in the total likelihood and resulted in the NLD and $\gamma$SF shown as blue circles in Fig. \ref{fig:nld_conv} and \ref{fig:gsf_conv}, respectively. 

\begin{table}
    \centering
    \caption{List of parameters used to normalize the NLD and $\gamma$SF. The spin-cut at $S_n$ $\sigma(S_n)$ is estimated from the model predictions of \cite{VonEgidy2005}, \cite{doi:10.1139/p65-139} and \cite{PhysRevC.80.054310} while the discrete levels spin-cut is estimated from the discrete states \cite{Capote2009} and shell model calculations \cite{Midtbo2018}. The s-wave resonance spacing $D_0$ are taken from \cite{Capote2009}, while the $\langle \Gamma_\gamma \rangle$ is a weighted average of tabulated radiative widths in \cite{PhysRevC.89.025810}.}
    \label{tab:param_table}
    \begin{tabular}{c c} \hline
    $S_n$ & $6.838$ MeV \\
    $D_0$ & $16.0(30)$ keV \\
    $\sigma(S_n)$ & $3.63(21)$ \\
    $E_d$ & $2.0$ MeV \\
    $\sigma(E_d)$ & $2.3(23)$ \\
    $\langle \Gamma_\gamma \rangle$ & $534(214)$ meV \\
    $\rho(S_n)$ & $1730(363)$ 1/MeV \\
    \hline
    \end{tabular}
\end{table}

\section{Discussion and comparison}

\subsection*{Level density}
We find that the experimental NLD fits exceptionally well with the tabulated discrete NLD up to about $E_x \approx 3.6$ MeV indicating that the level scheme might be complete up to even higher excitation energies, than the evaluated $E_x = 2.7$ MeV \cite{Capote2009}.

Comparing the two normalizations we see that the one including $\rho(S_n)$ results in a slightly steeper slope. Overall the two normalizations are well within the error-bars of each other demonstrating that normalization without knowledge of the NLD at the neutron separation energy are viable.

\begin{figure}
    \centering
    \includegraphics[width=0.5\textwidth]{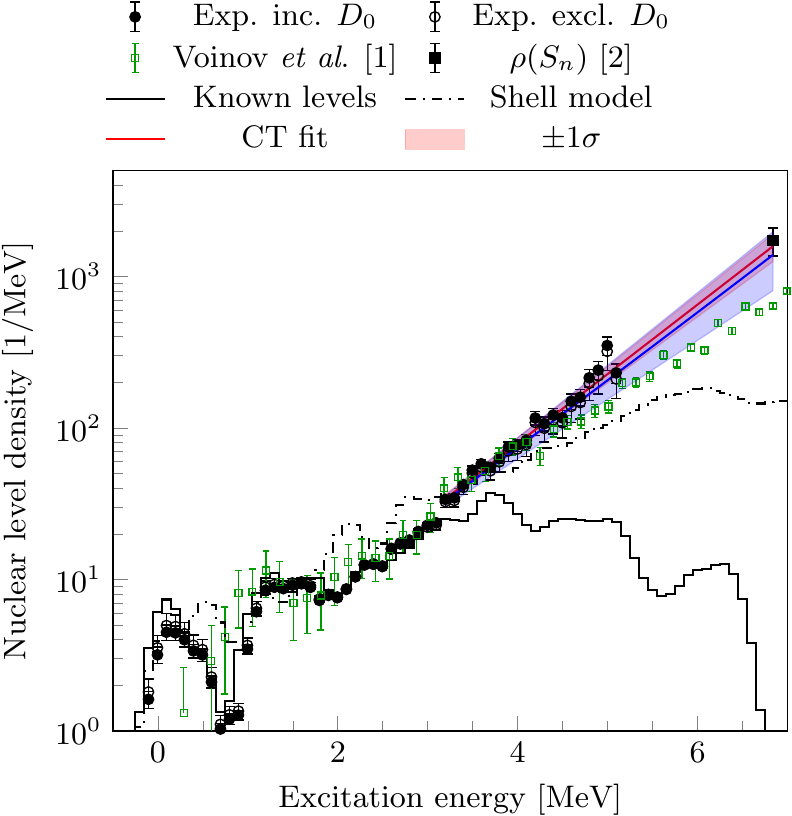}
    \caption{Comparison between the presented NLD shown by the open and filled black circles and the NLD found in large scale shell model calculations \cite{Midtbo2018} shown by the dash-dotted line. The green open boxes represent the NLD found in particle evaporation studies by A. Voinov \textit{et al}. \cite{Voinov2012}.}
    \label{fig:nld_comparison}
\end{figure}

Fig. \ref{fig:nld_comparison} shows the NLD compared with the experimental NLD found from particle evaporation spectra \cite{Voinov2012} and the NLD found in large scale shell model (SM) calculations \cite{Midtbo2018}. The SM results clearly overestimate the NLD between $1.8$ and $3.7$ MeV while underestimating above $4$ MeV up to around $6$ MeV where the model space seems to be exhausted.
The NLD found from evaporation studies fits well within the error bars up to about $4.5$ MeV where the presented NLD seems to tend to higher densities.

\begin{figure}
    \centering
    \includegraphics[width=0.5\textwidth]{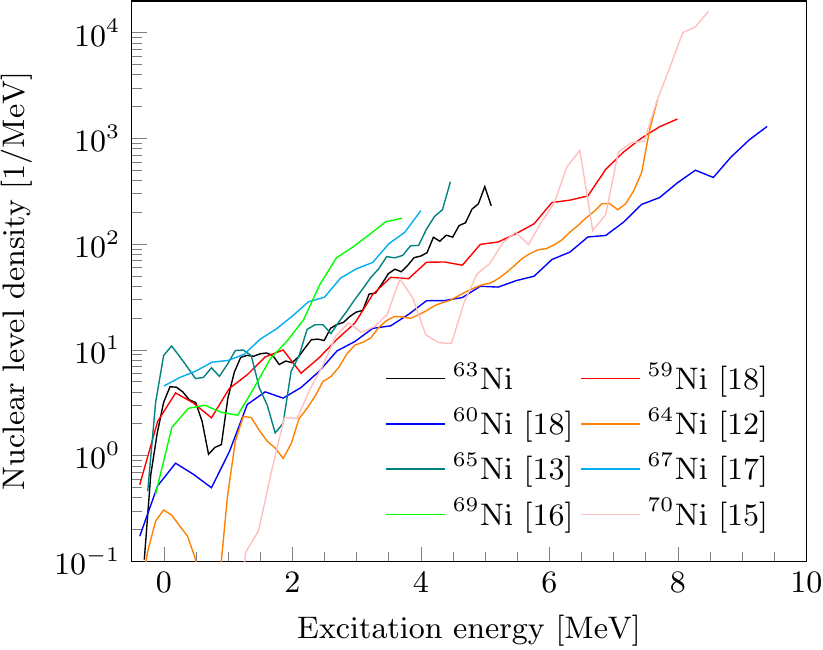}
    \caption{NLDs measured with the Oslo method in $\mathrm{Ni}$ isotopes.}
    \label{fig:nld_comp}
\end{figure}

In Fig. \ref{fig:nld_comp} the NLDs of $^{59,60,64,65,67,69,70}\mathrm{Ni}$ \cite{Renstrom2018,PhysRevC.94.044321,PhysRevC.96.014312,Ingeberg2022a,Spyrou2017,PhysRevC.97.054329} are shown together with the measured $^{63}\mathrm{Ni}$ isotope. We observe a clear trend with the absolute NLD increasingly with mass number while the temperature (i.e. the slope) decreases with mass number. 

\subsection*{\texorpdfstring{$\gamma$-ray strength function}{Gamma-ray strength function}}
The extracted $\gamma$SF features a strong upbend at low energies similar to what has been seen in other $\mathrm{Ni}$ isotopes \cite{Renstrom2018,PhysRevC.94.044321,PhysRevC.96.014312,PhysRevC.98.054619,PhysRevC.97.054329,Spyrou2017}, as well as other nuclei in the same mass region \cite{PhysRevLett.111.242504,Lar2017,PhysRevC.93.064302}. Comparing the measured strength function to the M1 strength predicted from the SM calculations in ref. \cite{Midtbo2018} we see that qualitatively these have a similar shape, although the absolute values of the SM calculations are considerably lower. Comparison with the photo-absorption cross section of $^{61}\mathrm{Ni}$ \cite{PhysRevC.98.054619} shows a reasonably good agreement as the giant dipole resonance evolves slowly with mass number. The normalized $\gamma$SF has a considerably large uncertainty band with the dominating contributing factor being the uncertainty in the average radiative width. Excluding the $\rho(S_n)$ in the normalization does also have a large impact on the uncertainties of the normalization for the $\gamma$SF, increasing the size of the error bars from $\approx 45\%$ to $\approx 80\%$, especially at higher $\gamma$-ray energies. 

\begin{figure}
    \centering
    \includegraphics[width=0.5\textwidth]{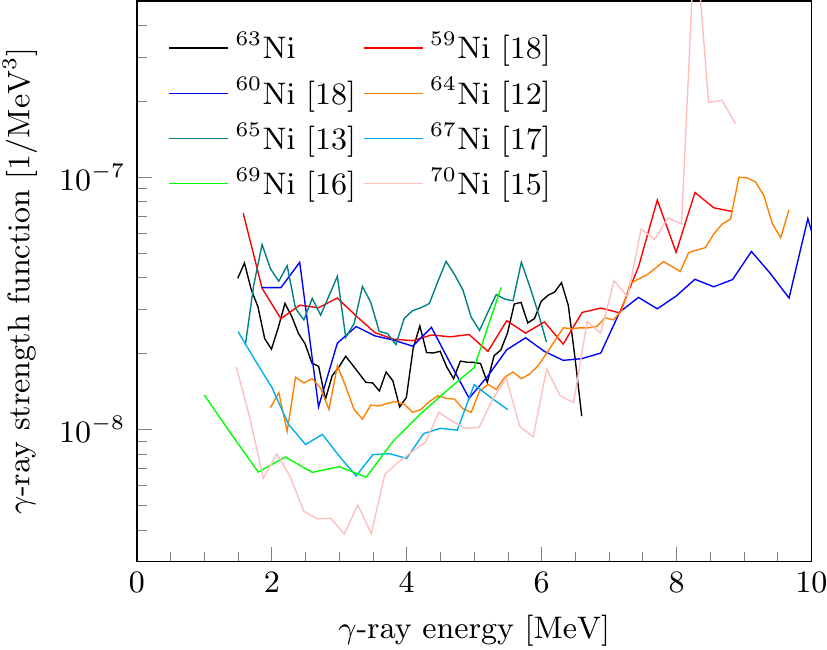}
    \caption{$\gamma$SFs measured with the Oslo method in $\mathrm{Ni}$ isotopes.}
    \label{fig:gsf_comp}
\end{figure}

In Fig. \ref{fig:gsf_comp} we show the $\gamma$SF for $^{59,60,64,65,67,69,70}\mathrm{Ni}$ \cite{PhysRevC.98.054619,PhysRevC.94.044321,PhysRevC.96.014312,Ingeberg2022a,Spyrou2017,PhysRevC.97.054329} together with the presented $\gamma$SF. From this comparison we can see a clear trend with the strength below $\approx 4.5$ MeV significantly decreasing with higher mass numbers. This is especially apparent in the unstable neutron rich nuclei ($A=67,69$ and $70$). The outlier are the $\gamma$SF of $^{65}\mathrm{Ni}$ which have the highest strength overall.

\section{Summary}
We have measured the NLD and $\gamma$SF of $^{63}\mathrm{Ni}$ and found that the NLD agrees well with that found from known levels, and are compatible with the NLD at the neutron separation energy found in neutron resonance studies. The NLD of \cite{Voinov2012} agrees with the presented NLD for excitation energies up to about $4.7$ MeV where the presented NLD seems to be somewhat steeper. Based on this we conclude that our results tend to favour the NLD found in resonance studies, rather than those of \cite{Voinov2012}.

The measured $\gamma$SF features a strong low energy enhancement similar to that found in other $\mathrm{Ni}$ isotopes. Shell model calculations from \cite{Midtbo2018} suggests that the enhancement may be due to M1 transitions within the quasi-continuum. Compared with $(\gamma,n)$ \cite{PhysRevC.98.054619} data for $^{61}\mathrm{Ni}$ there may be a pygmy resonance around $7$-$8$ MeV, but due to the large uncertainties in the absolute value of the measured $\gamma$SF we cannot conclude.

In general the exclusion of s-wave spacing in the overall fit of the NLD and $\gamma$SF resulted in very similar results, although with considerably larger uncertainties when extrapolating towards the neutron separation energy. Based on this we can conclude that if the level scheme is sufficiently well known a reasonably good normalization for the NLD can be obtained even without resonance data.

\begin{acknowledgments}
The authors would like to thank iThemba LABS operations for stable running conditions. This work is based on research supported by the Research Council of Norway under project grant no. 263030 (V.W.I, S.S.), by the National Research Foundation of South Africa under grants no. 105207, 99037, 90741 and 118846.
\end{acknowledgments}

\bibliography{references.bib}
\end{document}